# THE ETHICAL DILEMMA OF THE USA GOVERNMENT WIRETAPPING

Arwen Mullikin[1] and Syed (Shawon) M. Rahman[2]

[1]Graduate Student, Capella University
225 South 6th Street, 9th Floor Minneapolis, MN 55402, USA
*Email:* amullikin@email.capella.edu
[2]Assistant Professor of Computer Science University of Hawaii-Hilo, Hilo, HI, USA
and Adjunct Faculty, Capella University, Minneapolis, MN 55402,USA
*Email:* SRahman@Hawaii.edu

## ABSTRACT

*USA Government wiretapping activities is a very controversial issue. Undoubtedly this technology can assist law enforced authority to detect / identify unlawful or hostile activities; however, this task raises severe privacy concerns. In this paper, we have discussed this complex information technology issue of governmental wiretapping and how it effects both public and private liberties. Legislation has had a major impact on the uses and the stigma of wiretapping for the war on terrorism. This paper also analyzes the ethical and legal concerns inherent when discussing the benefits and concerns of wiretapping. The analysis has concluded with the effects of wiretapping laws as they relate to future government actions in their fight against terrorists.*

## 1. INTRODUCTION

We believe the advances of technology allows for amazing advances which have kept our the United States safe from terrorists for the last eight years. The ethical sides of our practices have caught national attention and a firestorm of debates has risen on wiretapping and the ethical concerns, as well as, the related laws. "Developments in technology have also had a profound impact on privacy. To attempt to function in modern society without employing telecommunications is to be eccentric"[1].

There is no doubt that wiretapping is very controversial and causes deep passions to arise from those who are for wiretapping and for those who are against it. In 2009 we observed the United States District Judge Vaughn Walker threaten to side in favor of the plaintiffs if the Obama Administration did not produce documents showing communications intercepted without warrants against Al-Haramain Islamic Foundation officials [2]. This wiretapping case has been going on for three years and there does not seem to be a resolution in site.

Wiretapping is not new to presidents, in fact it has been going on since the 1800's, just not as high tech as it is today. The United States Constitution gives the president power to protect the country from national attacks. And who would not want that? The ethical dilemma lies in not just the legality of wiretapping, but in the question of whether our private and public freedoms are being run over by a zealous government, all in the name of national security.

The United States Constitution does not give citizens an inalienable right to privacy but the Fourth Amendment does protect US Citizens from government illegal searches and seizures. It is a balancing act between protecting US Citizens private and public freedoms, and protecting their liberties from being trounced upon. Wiretapping has been around since the 1800s, and is every bit as controversial as it was back then. The history of wiretapping will clearly show the controversial pros and cons of this issue.

## 2.0 HISTORY OF WIRETAPPING

Technology such as wiretapping has enabled government to reach past borders to protect its citizens like never imagined before. Rasch [3] stated, "Whenever a new technology is developed, or a new threat that causes us to deploy these technologies, questions invariably arise about their legality". The government has participated in wiretapping during the 1960s and 1970s when they implemented a widespread program of wiretapping and permeation of dangerous groups thought to be an endangerment to the government. This program was fraught with abuses and even spied on Dr. Martin Luther King, Jr. The first step in analyzing wiretapping is first to go back to the beginning and look at what the United States Constitution says, since the Constitution is the law of the land in the United States of America.

"The right of the people to be secure in their persons, houses, papers, and effects, against *unreasonable searches and seizures* [Italics added for emphasis], shall not be violated, and no Warrants shall issue, But upon probable cause, supported by Oath or affirmation, and particularly describing the place to be searched, and the persons or things to be seized" (The Fourth Amendment of the U.S. United States Constitution).

Warrantless wiretapping has been sanctioned by the Supreme Court and the full case (UNITED STATES V. UNITED STATES DIST. CT., 407 U. S. 297 (1972) was posted on the U.S. Supreme Court website, www.supreme.Justia.com [4], the law suit states that many presidents and attorney generals since 1946 and the Executive branch of government have used warrantless electronic surveillance to collect intelligence data since the 1800s. Wiretapping was instituted to speed up the process and to avoid the red tape of bureaucracy. As with all things government, the pace in which things are done is extremely slow. Wiretapping would circumvent the process and allow quick actions to be taken to apprehend the enemy.

## 3.0 PRIVACY ISSUE

The meaning of privacy has changed throughout the years but has come to be understood as freedom from having one's personal life and information tampered with. Nowhere in the United States Constitution or the Bill of Rights do we see privacy as an inalienable right. Some scholars and law experts use the Fourth Amendment to infer that all citizens have this right due to the search and seizure section. The case of Griswold v. Connecticut is often used as the defining case where the Supreme Court ruled that we, as individuals, have a right to privacy.

In today's technology age we see personal privacy compromised all the time through unsolicited cell phone pictures and videos being taken without consent, identity theft, phishing, spamming, spoofing, and even companies using our web browser cookies from Internet sites we access to gain more information about how we shop. Citizens believe, for the most part, that there are situations that limit their privacy in a high technology age; however, even then most citizens believe they normally still have some control over the lack of privacy in our technology age. Things like turning off cookies, prosecuting identity thieves, and using a number of blocking tools and technologies to protect our actions. But there is one area where most citizens feel they cannot turn off or even control cyber-technology and that is when it comes to government surveillance.

Most of the people who live in the United States want to believe that our government will do no harm and that they try there to stop terrorists through checks and balances, laws, and congress. So the ethical dilemma then becomes did the government do harm by "sneak and peek" on United States citizens through the use of wiretapping? Some people would shout a resounding YES while others would say the government has the power to protect the citizens, as we will see next section.

## 3.1 WHAT ARE THE SIDES OF THE ETHICAL DILEMMA

On one side of the argument we have a loss privacy of citizen's, which is inevitable and necessary, if we are to protect our country from terrorist attacks. On the other side, our civil liberty groups and others will argue that giving up basic privacy right, which are the bedrock of our government, is a mighty high price to pay for being protected.

The argument against government wiretapping is, if government wiretapping is allowed then the government will push the boundaries and try to get as much power as possible. Taking control over every aspect of our lives and implementing total rule, while doing away with the United States Constitution and the America we know today.  This group always says "This blatant abuse of government power obviously cannot be permitted and must be stopped while we still have rights".

Opponents against wiretapping argue that government is too big and should be controlled and limited in what they can do. This is achieved by due process of rights and prevents misuse and abuse of governmental powers. These critics point out that during the 60s and 70s emancipated power was misused and so it is by today's government[5].

The main uneasiness behind wiretapping is that the government collects, and keeps forever, a large amount of information about individuals in the U.S., including citizens. Are we moving to a government that maintains our DNA and knows everything about us? This is ultimately the under lying fear. Out of this fear comes the dilemma of how to keep American citizens safe, while at the same time, not having government over step their boundaries of individual privacy, as granted by the United States Constitution. The question remains, by those supporting this argument, is can the government be trusted to not abuse or misuse power given to them? Can the government be trusted to do the right thing when they have unrestrained power? If history is any indication, the answer leans towards no.

John Yoo, former Justice Department [6] of the USA, stated "It was instantly clear after Sept. 11, 2001, that our security agencies knew little about al Qaeda's inner workings, could not detect its operatives' entry into the country, nor predict where it might strike next". Yoo went on to further to say that wiretapping is the only way to detect follow-up attacks "and what president -- of either political party -- wouldn't immediately order the National Security Agency (NSA) to start, so as to find and stop the attackers"?

Foreign Intelligence Surveillance Act (FISA) was created after the Cold War and was never meant to squash or hinder military operations. FISA is an outdated law that cannot address terrorist attacks. FISA proved unable to protect America from the 9/11 terrorist attacks, which were carried out without any prevention from FISA. It is imperative to consistently monitor terrorist communications and all of their communication channels. Through the use of governmental wiretapping many terrorists have been caught since 9/11 and our country has been protected from terrorists and any major attacks. On the other side, are the opponents in favor of governmental wiretapping. These proponents believe the government should be able to use any means necessary to protect America from terrorism.

They believe it is the president's job and responsibility to protect the security and interest of the people of this country. Congress is too numerous to act quickly to live threats, and in war, it is imperative that the president be unhindered in carrying out his executive orders by allowing wiretapping. With that said, those who favor wiretapping can look at the issue of privacy and be at ease with loosing some of their privacy for the greater good of the country.

## 3.2 IN FAVOR OF WIRETAPPING

The United States laws, regarding wiretapping, appears to be on the side of the United States government. The government relies on the following laws for support of wiretapping.

(i) The first law is the Authorization for Use of Military Force (AUMF) passed September 18, 2001. In this law there is a line that gives the president power to use all necessary and appropriate force against nations, organizations, or persons that he deems as ones who have planned, authorized, committed, or aided in the September 11, 2001 terrorist attacks. In most cases warrants for wiretapping are issued under the Electronic Communications Privacy Act (EC). We must have probable cause that a crime has been or will be committed and that the wiretap will unveil evidence of that crime. This act allowed for traceable devices to be added to phones of suspects and also show Internet sites visited.

(ii) The next law to be passed in support of wiretapping was the USA-Patriot Act [7]. The Patriot Act authorized roving wiretaps which allowed more power to the FBI and circumvent of the courts. The Patriot Act still upholds the EC and wiretapping can only be done in regards to criminal activity. Since most terrorist commit criminal activities the law then allows for wiretapping.

(iii) The Foreign Intelligence Surveillance Act (FISA) which, Cornell University [8] describes as the "procedures for requesting judicial authorization for electronic surveillance and physical search of persons engaged in espionage or international terrorism against the United States on behalf of a foreign power". FISA allows for the executive branch to obtain an interception or seizure order by proving to a secret Foreign Intelligence Surveillance court that the surveillance will produce foreign intelligence regarding terrorism. What is unique about FISA is that it is directed at U.S. Citizens, permanent resident aliens, and corporations. In 1994 FISA was amended by Congress to include covert physical entries as well as electronic eavesdropping and wiretapping. In 1998 FISA was amended again by Congress to permit pen/trap orders, which, record telephone numbers and business records. If a US citizen is involved or suspected then probable cause of espionage must be shown. Later through the Patriot Act the FISA court was bypassed altogether.

(iv) The Homeland Security Act 2002 gave increased powers to law enforcement agencies to track down suspected terrorist and criminals. This act allowed an increase in monitoring of e-mail and cell phone communications [9]. In 2005 the Bush Administration argued that the Patriot Act did not overstep the Executive

Powers but that this was needed to protect America. In 2009 the Obama Administration defended the Bush wiretapping against the Electronic Frontier Foundation (EFF) in San Francisco[10].

(v)  The Protect America Act is another law that would modernize FISA and protect Americans from terrorist attacks. The Department of Justice [11] had this to say about the Protect America Act:

> The Protect America Act modernized the Foreign Intelligence Surveillance Act (FISA) to provide our intelligence community essential tools to acquire important information about terrorists who want to harm America. The Act, which passed with bipartisan support in the House and Senate and was signed into law by President Bush on August 5, 2007, restores FISA to its original focus of protecting the rights of persons in the United States, while not acting as an obstacle to gathering foreign intelligence on targets located in foreign countries. By enabling our intelligence community to close a critical intelligence gap that existed before the Act became law, the Protect America Act has already made our Nation safer.

The bottom line, as believed by those in favor of wiretapping, is that the United States Constitution has bestowed powers to the president to protect this country from any nation or individual group from carrying out an attack on it. Gathering intelligence of all kinds is a key part of war. The United States has declared war on terrorism, intelligence gathering should not be different from any other time of war. Our military and intelligence agencies must know where the target is, without wiretapping they would be shooting in the dark.

In 2009 the proponents of wiretapping won a step forward when a rare public ruling from a secret federal appeals court said "telecommunications companies must cooperate with the government to intercept international phone calls and e-mail of American citizens suspected of being spies or terrorists [11].

People in favor of wiretapping believe that wiretapping is absolutely ethical and should be used by intelligence agencies to protect the citizens of the United States from those who would do us harm. The war on terror is far different than any other war America has ever fought, the rules are fuzzy, the enemy is slippery and fast, and our very homeland has become the battle ground and a place to hide and plan attacks. For this reason many people believe the Government needs the power to wiretap. Now we will take a closer look at the argument for the other side that says Government has overstepped their boundaries of privacy.

## 3.3 OPPOSITION TO WIRETAPPING

There are three laws that make it illegal to engage in wiretapping: 50 ISC 1809(a), 47 USC 605, and 18 USC 2511. These laws make wiretapping legal only through consent of one or all parties, or by consent of the telecommunications service provider, and finally by court order [12]

(i)  Due to the Nixon administration's illegal use of wiretapping of subversive groups, Congress passed the FISA. FISA scales back presidential abuse in conducting foreign intelligence investigations. The FISA law states that FISA now "shall be the exclusive means by which electronic surveillance and the interception of domestic wire, oral, and electronic communications may be conducted (Section 201 of the FISA as enacted in 1978)".

(ii) With the passing of the Patriot Act many people saw the government overstepping their powers particularly with section 215 (Patriot Act), which, gives the FBI authority to obtain library, and bookstore receipts of individuals. This section also imposes a gag order, which forbids those who provided the information to inform the party of interest. Section 215 gives the FBI leeway to search any records it believes pertinent to terrorism investigations, even though those people may not be suspected of criminal conduct. More than 200 search orders have been issued, since 2003, by the Department of Justice [13]. The American Library Association believes that the Patriot Act violates people's First Amendment rights to freedom of Speech by not being able to notify the parties that the information pertains to.

In 2005 wiretapping really came to the forefront and became a major issue in America. During President Bush's Administration, the government was involved in domestic spying on citizens through capturing emails and phone calls to those made outside of the United States. Congress stated this was far beyond the boundaries of FISA and that domestic spying on citizens was not allowed. Wiretapping was seen as such a violation of privacy for citizens that the Electronic Frontier Foundation (EFF) sued the government and officials who implemented the program in September 2009 in an effort to get the government to stop using wiretapping [14].

The ethical dilemma question then become "is it right for the government to use domestic spying on its citizens"? Those who oppose wiretapping would say a resounding NO. Their argument lies in the fact that there is never a time that the government should spy on its own citizens without a search warrant. The search warrant and laws protect the citizens from a government that has swelled with power and takes liberties they should not be taking.

Opponents against wiretapping would point to the government's recent activities to show that they have already taken excessive power and are looking for more. This can be seen in the recently passed legislations regarding: Net Neutrality, the Fairness Doctrine, and S 773.3 Cyber Security Act 2009 bill. All of these legislations aim at giving more control to the government. The government has also shown, through President Nixon's actions that absolute power corrupts absolutely. Opponents against will also say that even the good get corrupted with no restraints, therefore, warrants, monitors and laws keep the government from intruding into our lives excessively.

# 4.0 THE EFFECTS OF WIRETAPPING LAWS FOR FUTURE GOVERNMENT ENDEAVORS

In the ongoing debate we do know that each side wins a little every couple of years to keep the battle going. Michael Hewitt [14] stated:

The role of wiretapping as given by the powers invested in the Patriot Act, particularly the United States Constitutionality of this method, includes what criteria must be met to preserve United States Constitutionally protected civil liberties. Wiretapping has had a significant effect, as given by the powers of the Patriot Act, on both the personal security and privacy of the American people. Current wiretapping policy lacks clear and appropriate guidelines addressing the modern terrorist threat. Future policy should reflect the need for new criteria to protect the national security and respect United States Constitutionally imposed limits.

So the question then we need to ask as a nation is "Will the USA become like a science fiction movie, where small metal objects are floating above us and recording our every move and word"? As far as we can see today that is highly unlikely and left for Hollywood to create.

History has shown us that wiretapping is not going away anytime soon and does have a place in government when it comes to protecting the citizens of the United States of America. In 2006 we saw that President Bush was successful at blocking a Justice Department Investigation on the National Security Agency's (NSA) wiretapping program. The president made the decision based on what he felt was more important to protect, the secrecy and security of the wiretapping program and limiting the number of persons involved in order to not compromise security [15].

In July 2009 we have observed that the Obama Administration has defended the practice of the Bush Administration wiretapping program in San Francisco in a lawsuit with EFF [16]. Though no details of this case were released to the public the Inspector General's Office mentioned that the wiretapping program was useful in defending America.

## 5.0 CONCLUSION

In conclusion, the fight for and against wiretapping will not be going away any time soon as both sides battle for what they believe is right. The first side battles to protect American from being attacked and believes the government has the right to use whatever means necessary to accomplish that job. The second side believes that American freedoms are so important that the government should only do what is in the United States Constitution and nothing more. This side believes that it is a slippery slope to start allowing the government to take over any of those rights, even if it is to protect a nation.

The truth lies somewhere in the middle. The government should have the power to protect the citizen's of this great land but they must balance that responsibility delicately and be ever mindful of the rights of the citizens they are protecting.

How this debate will end know one can truly say. We can be assured more cases will be brought before our judges to decide this matter. America also knows that the security of the United States and all its freedoms do not come without a price, the war on terrorism has shown that very clearly.

## Authors

**Arwen Mullikin**

Ms. Arwen Mullikin is a graduate student in the Information and Assurance program at Capella University. She has over 15 years in the technology field and has held positions as a Unix System Administrator and Project Manager. She looks forward to graduating December 2010 and starting her doctorate degree next year.

**Syed (Shawon) M. Rahman**

Syed (Shawon) Rahman is an Assistant Professor in the Department of Computer Science & Engineering at the University of Hawaii-Hilo and a adjunct faculty of Information Technology, Information Assurance & Security at the Capella University. Dr. Rahman's research interests include Software Engineering Education, Data Visualization, Data Modelling, Information Assurance & Security, Web Accessibility, and Software Testing & Quality Assurance. He has published more than 50 peer-reviewed papers. He is an active member of many professional organizations including ACM, ASEE, ASQ, IEEE, and UPE.